\PassOptionsToPackage{table}{xcolor}
\documentclass[sigconf]{acmart}
\AtBeginDocument{%
  }

\setcopyright{acmlicensed}
\copyrightyear{2018}
\acmYear{2018}
\acmDOI{XXXXXXX.XXXXXXX}
\acmConference[Conference acronym 'XX]{Make sure to enter the correct
  conference title from your rights confirmation email}{June 03--05,
  2018}{Woodstock, NY}
\acmISBN{978-1-4503-XXXX-X/2018/06}

\usepackage{amssymb}
\usepackage{array}
\usepackage{tcolorbox}

\newcommand{\cross}{$\times$} 

\definecolor{lightgreen}{RGB}{204, 255, 204} 
\definecolor{lightyellow}{RGB}{255, 255, 204}
\definecolor{lightpurple}{RGB}{230, 230, 250}

\newtcbox{\hlgreen}{on line, colback=lightgreen, boxrule=0pt, arc=5pt, left=0pt, right=0pt, top=0pt, bottom=0pt}

\newtcbox{\hlyellow}{on line, colback=lightyellow, boxrule=0pt, arc=5pt, left=0pt, right=0pt, top=0pt, bottom=0pt}

\newtcbox{\hlpurple}{on line, colback=lightpurple, boxrule=0pt, arc=5pt, left=0pt, right=0pt, top=0pt, bottom=0pt}

\begin{document}

\title{DaemonSec:
Examining the Role of Machine Learning for Daemon Security in Linux Environments}

\author{Sheikh Muhammad Farjad}
\affiliation{%
  \institution{University of Nebraska at Omaha}
  \city{Omaha}
  \state{Nebraska}
  \country{USA}}
\email{sfarjad@unomaha.edu}

\renewcommand{\shortauthors}{Farjad et al.}

\begin{abstract} % not more than 200. if possible: preferably closer to 150
\texttt{DaemonSec} is an early-stage startup exploring machine learning (ML)-based security for Linux daemons, a critical yet often overlooked attack surface. While daemon security remains underexplored, conventional defenses struggle against adaptive threats and zero-day exploits. To assess the perspectives of IT professionals on ML-driven daemon protection, a systematic interview study based on semi-structured interviews was conducted with 22 professionals from industry and academia. The study evaluates adoption, feasibility, and trust in ML-based security solutions. While participants recognized the potential of ML for real-time anomaly detection, findings reveal skepticism toward full automation, limited security awareness among non-security roles, and concerns about patching delays creating attack windows. This paper presents the methods, key findings, and implications for advancing ML-driven daemon security in industry.
\end{abstract}




\maketitle
%---------------------------------------------------------%
\section{Introduction and Motivation} \label{sec:introduction}
%---------------------------------------------------------%

Linux-based systems are central to servers, cloud environments, mobile devices, and edge computing. With a vast codebase of around thirty million lines, the Linux kernel and its distributions are susceptible to vulnerabilities~\cite{codebase}. These vulnerabilities can have severe consequences for organizations and individuals relying on Linux systems. Therefore, the security of Linux-based systems is of paramount importance and must be considered seriously.

Although the open-source nature of the Linux ecosystem fosters innovation and collaboration, it also introduces security risks arising from the exploitation of vulnerabilities in third-party dependencies. In recent years, supply chain attacks have become a major concern, allowing attackers to inject vulnerabilities into widely used open-source components~\cite{opensource}. Despite active security monitoring within the Linux community, the decentralized nature of open-source software development makes it challenging to prevent such intrusions effectively~\cite{opensource2, kubernetes}. While Linux security has been extensively studied, limited research specifically focuses on daemons, creating a crucial gap in proactive defense strategies~\cite{daemonsecurity}. Furthermore, zero-day attacks and the increasingly complex infrastructure of Linux systems render traditional signature-based security measures ineffective. Emerging machine learning (ML) approaches have shown promise for real-time anomaly detection and intrusion prevention. However, their application to daemon security remains largely unexplored, allowing daemon-focused attacks to continue. This gap led to the research question: \textit{\textbf{How do IT professionals in industry and academia perceive the issue of daemon security, and what are their perspectives on mitigating it through an ML-based approach?}}

To address this question, a systematic interview study was conducted with 22 IT professionals from both industry and academia. This paper presents the findings and discusses their implications for advanced Linux security frameworks, including the \texttt{DaemonSec} startup~\cite{daemonsec}.

%---------------------------------------------------------%
\section{Background and Related Work} \label{sec:background}
%---------------------------------------------------------%
Daemons are privileged background processes in Linux systems responsible for managing essential tasks such as networking, authentication, and system monitoring~\cite{stallings2018}. Their continuous execution and elevated privileges make them prime targets for cyberattacks, including privilege escalation, persistence mechanisms, and remote exploitation. Recent ransomware campaigns have exploited supply chain vulnerabilities in background services (i.e., daemons) to infiltrate critical infrastructure, underscoring the urgent need for stronger security mechanisms~\cite{ransomware}.

Existing Linux security research has primarily focused on kernel hardening~\cite{kernel-hardening}, access control mechanisms~\cite{SELinux}, and general intrusion detection systems~\cite{intrusion-detection}. However, studies specifically addressing daemon security remain limited. Traditional signature-based detection systems struggle against zero-day exploits and polymorphic malware, leading to growing interest in ML-based anomaly detection~\cite{zero-day, farjad2020}. Prior research has demonstrated the potential of ML for security applications, such as intrusion detection and malware classification, yet its application to daemon security remains in its early stages~\cite{dos, intrusion-detection}.

One of the key challenges in adopting ML-based security solutions is industry skepticism, driven by various factors. For instance, Arp et al.~\cite{dos} identify common pitfalls in ML-oriented security research (e.g., sampling bias, data snooping, and spurious correlations), which can deter practitioners from deploying ML-based solutions in production environments. This study contributes to the ongoing discourse by examining how IT professionals perceive daemon security risks and the feasibility of ML-driven security approaches.

%---------------------------------------------------------%
\section{Methods: Hypotheses and Study Design} % Study Design
%---------------------------------------------------------%

\subsection{Hypotheses}
The following hypotheses were formulated to answer the research question developed in Section~\ref{sec:introduction}:

\begin{itemize}
    \item \textbf{H1:} Daemon security is an underexplored area within the cybersecurity domain.
    \item \textbf{H2: }ML-based solutions are preferred over traditional security methods for daemon protection.
\end{itemize}

\subsection{Ecosystem Model}
\begin{figure}[h]
    \centering
    \includegraphics[width=0.47\textwidth]{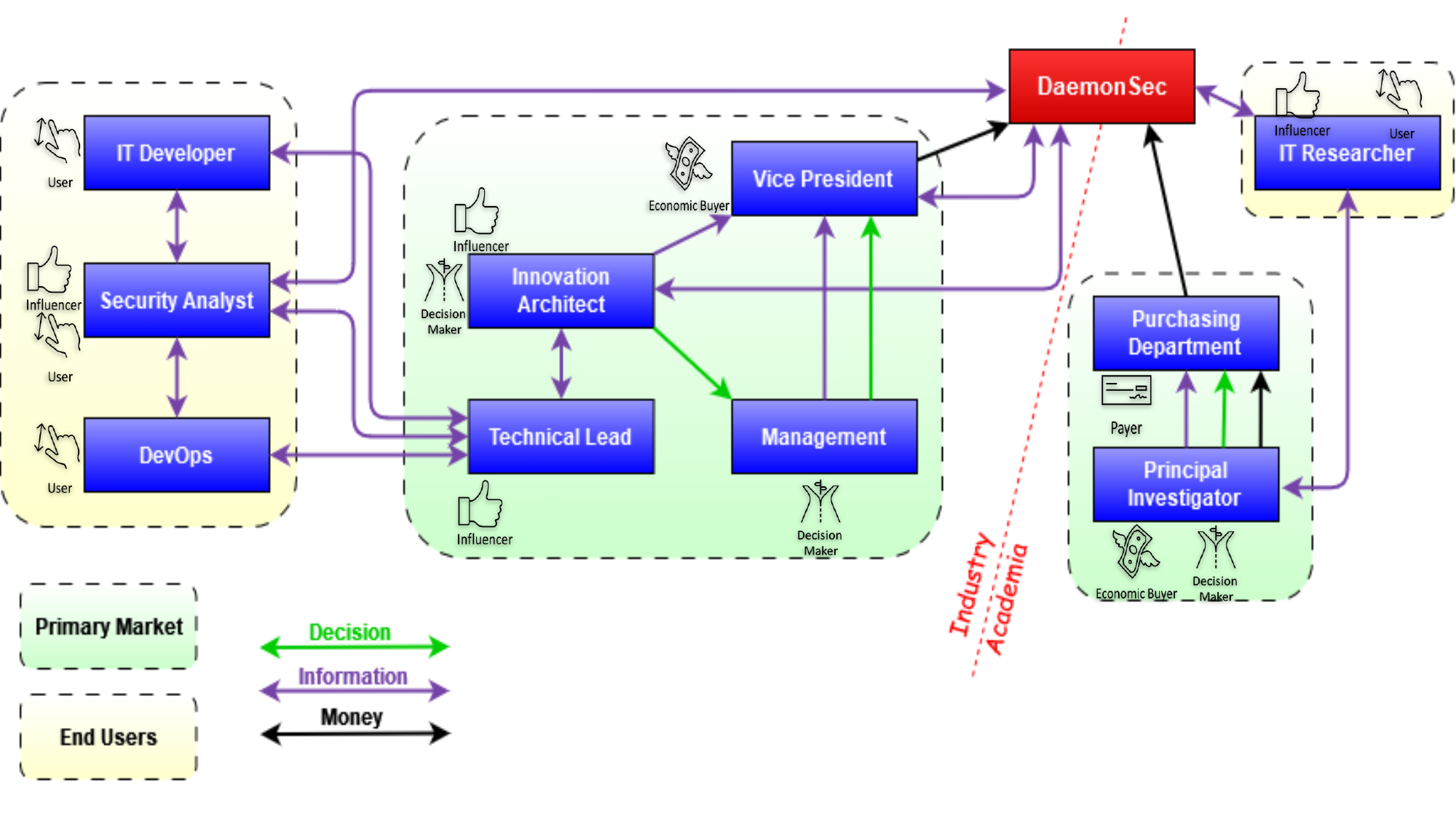}
    \caption{Customer Ecosystem Model of DaemonSec}
    \label{fig_ecosystem}
\end{figure}

To capture all relevant stakeholders, the customer-centric aspects of the business model canvas framework were used to develop the customer ecosystem model of \texttt{DaemonSec}~\cite{BMC2, BMC}. The model, illustrated in Figure~\ref{fig_ecosystem}, accounts for the hierarchical structure of stakeholders, which assigns greater value to higher-level roles in the primary market. For example, feedback from executives (e.g., vice presidents and innovation architects) carries more weight than that of IT developers or DevOps professionals, reflecting their decision-making authority in security adoption.

\subsection{Customer Discovery Interviews}
The study included 22 participants: 18 males (81.8\%) and 4 females (18.2\%), with 19 from the United States and three from Australia, Saudi Arabia, and Pakistan. Thematic saturation was reached at 22 interviews, prompting the conclusion of data collection. Based on the customer ecosystem model (Figure~\ref{fig_ecosystem}), the majority of participants (n=17) were from industry, while five represented academia.

Of the 22 participants, 21 were interviewed in real time \textit{via} Zoom video conferencing, with recordings made for annotation and analysis. One participant, due to scheduling constraints, was interviewed \textit{via} their official email. The interviews were conducted in three iterative rounds, allowing for progressive refinement of questions based on prior responses. The questions were designed to elicit relevant insights while minimizing interviewer bias. Ethical considerations, including informed consent and participant confidentiality, were strictly adhered to in compliance with research guidelines.

%---------------------------------------------------------%
\section{Results and Discussion}
This section presents key findings from customer discovery interviews on hypothesis validation and security insights. 

\subsection{Hypothesis Validation}

Both hypotheses were supported by participant responses:

\begin{itemize} 
    \item \textbf{H1} (Daemon security is underexplored) – 77.27\% of participants were unaware of daemons and, despite working in IT, had no experience managing background services. Thus, H1 is validated. 
    \item \textbf{H2} (ML-based security is preferred) – 95.45\% of participants favored ML-driven models; however, concerns about automation risks and trust issues were also noted. These findings suggest the need for a hybrid approach that integrates ML-based security with traditional signature-based systems. Thus, H2 is validated and offers additional insights.
\end{itemize}

\subsection{Key Security Insights} 
Semi-structured interviews facilitated in-depth discussions and elicited participants' perceptions, which not only contributed to hypothesis testing but also provided key insights relevant to startups focused on ML-based security.

\textbf{Skepticism toward full automation.}
Participants expressed concerns about false positives and the lack of human oversight, favoring hybrid security models over fully automated solutions.

\textbf{Limited security awareness among non-security stakeholders.}
Developers and DevOps teams often rely on dedicated security teams, resulting in reactive rather than proactive security practices. This reliance also absolves them of the need to invest effort in understanding and implementing security measures.

\textbf{Delays in security patching create attack windows.}
A participant with advanced knowledge of Linux systems highlighted that time gaps between patch release and deployment provide adversaries with opportunities for exploitation, particularly in high-privilege daemons.

%---------------------------------------------------------%
\section{Concluding Remarks}

This study confirms that daemon security is a less-explored domain, with 77.27\% of participants unaware of daemons and their operations. While participants favored ML-based security measures over traditional methods, some reluctance remained regarding their exclusive adoption, primarily due to a lack of trust. Therefore, a hybrid security solution combining ML methods with traditional signature-based approaches is preferred by the existing customer ecosystem. This central takeaway, along with other key insights from the interviews, will inform the future design of \texttt{DaemonSec}.

%---------------------------------------------------------%

\bibliographystyle{ACM-Reference-Format}
\bibliography{sample-base}

\appendix

\section{Thematic Flow of Interview Questions}
 
\begin{figure}[H]
    \centering
    \includegraphics[width=0.3\textwidth]{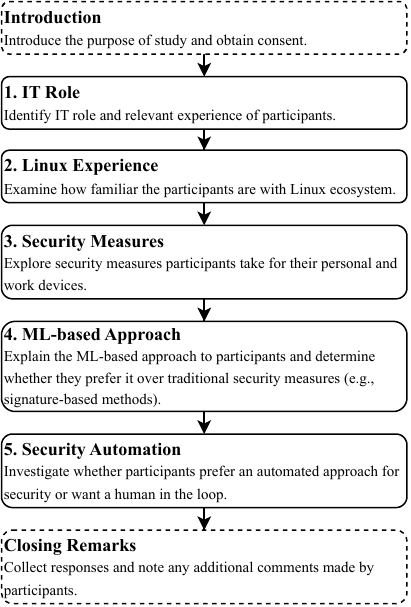}
    \caption{Overview of Thematic Flow in Interview Questions}
    \label{fig_interview}
\end{figure}

Figure~\ref{fig_interview} provides an overview of the thematic flow of the interview questions, which were asked to the participants in the illustrated sequential order.

\section{Interview Summary}
Table~\ref{tab:test_hypotheses} provides a summary of 22 interviews conducted in three iterations: \hlgreen{Round \#1}, \hlyellow{Round \#2}, and \hlpurple{Round \#3}. Each record represents an individual participant and the outcomes of their responses, i.e., whether they validate (\checkmark) or invalidate (\cross) the hypotheses (Hypo. 1 and Hypo. 2).

\begin{table}[H]
  \caption{Summary of Interviews}
  \label{tab:test_hypotheses}
  \centering
  \small % Adjust font size
  \begin{tabular}{llllcc}
    \toprule
    \textbf{ID} &\textbf{Gender} & \textbf{Country} & \textbf{Segment} & \textbf{Hypo. 1} & \textbf{Hypo. 2}   \\
    \midrule
    \rowcolor{lightgreen} P1 &Male &USA & Academia & \checkmark & \checkmark \\
    \rowcolor{lightgreen} P2 &Male &USA & Academia & \checkmark & \checkmark \\
    \rowcolor{lightgreen} P3 &Female &Australia & Industry & \checkmark & \checkmark \\
    \rowcolor{lightgreen} P4 &Male &USA & Industry & \checkmark & \checkmark \\
    \rowcolor{lightgreen} P5 &Female &Pakistan & Industry & \checkmark & \checkmark \\
    \rowcolor{lightgreen} P6 &Male &USA & Industry & \cross & \checkmark \\
    \rowcolor{lightgreen} P7 &Male &USA & Industry & \checkmark & \checkmark \\
    \rowcolor{lightgreen} P8 &Male &USA & Academia & \checkmark & \cross \\
    \rowcolor{lightgreen} P9 &Male &USA & Industry & \checkmark & \checkmark \\
    \rowcolor{lightgreen} P10 &Male &USA & Industry & \checkmark & \checkmark \\
    \rowcolor{lightyellow} P11 &Male &USA & Industry & \cross & \checkmark \\
    \rowcolor{lightyellow} P12 &Male &Saudi Arabia & Industry & \cross & \checkmark \\
    \rowcolor{lightyellow} P13 &Male &USA & Industry & \checkmark & \checkmark \\
    \rowcolor{lightyellow} P14 &Male &USA & Industry & \checkmark & \checkmark \\
    \rowcolor{lightyellow} P15 &Female &USA & Academia & \checkmark & \checkmark \\
    \rowcolor{lightpurple} P16 &Male &USA & Industry & \checkmark & \checkmark \\
    \rowcolor{lightpurple} P17 &Male &USA & Industry & \cross & \checkmark \\
    \rowcolor{lightpurple} P18 &Male &USA & Industry & \checkmark & \checkmark \\
    \rowcolor{lightpurple} P19 &Female &USA & Academia & \checkmark & \checkmark \\
    \rowcolor{lightpurple} P20 &Male &USA & Industry & \checkmark & \checkmark \\
    \rowcolor{lightpurple} P21 &Male &USA & Industry & \cross & \checkmark \\
    \rowcolor{lightpurple} P22 &Male &USA & Industry & \checkmark & \checkmark \\
    \bottomrule
  \end{tabular}
\end{table}

\end{document}